\RequirePackage{arydshln}
\documentclass[aps,twocolumn,nofootinbib,superscriptaddress,preprintnumbers,pra,10pt,floatfix]{revtex4-1}

\usepackage[utf8]{inputenc}
\usepackage{float}
\usepackage{colortbl}
\usepackage{amsmath,amssymb,amsfonts, bm,bbm,slashed}
\usepackage{dsfont} 
\usepackage{hyperref}
\usepackage{graphicx}
\usepackage{enumitem}
\usepackage{arydshln}
\usepackage{mathtools}
\usepackage{bbold}
\usepackage{braket}
\usepackage{soul}

\usepackage{feynmp-auto}
\makeatletter
\g@addto@macro\bfseries{\boldmath}
\makeatother

\newcommand{\be} {\begin{equation}}
\newcommand{\ee} {\end{equation}}
\newcommand{\bea} {\begin{eqnarray}}
\newcommand{\eea} {\end{eqnarray}}
\newcommand{\no} {\nonumber}
\newcommand{\ba} {\begin{array}}
\newcommand{\ea} {\end{array}}
\newcommand{\gsim}{\lower.7ex\hbox{$\;\stackrel{\textstyle>}{\sim}\;$}}
\newcommand{\lsim}{\lower.7ex\hbox{$\;\stackrel{\textstyle<}{\sim}\;$}}

\newcommand{\msbar}{$\overline{\mathrm{MS}}$}

\newcommand{\cL}{\mathcal{L}}
\newcommand{\cC}{\mathcal{C}}
\newcommand{\cO}{\mathcal{O}}
\newcommand{\cA}{\mathcal{A}}

\newcommand{\xG}{x_{G^\prime}}

\newcommand{\mZ}{m_{Z^\prime}}

\newcommand{\Gp}{{G^\prime}}
\newcommand{\Zp}{{Z^\prime}}
\newcommand{\us}{m_U^2}
\newcommand{\gp}{\hat{g}_4}
\newcommand{\epsIR}{\epsilon_{\rm IR}}
\newcommand{\epsUV}{\epsilon_{\rm UV}}
\newcommand{\LmuIR}{L_\mu^{\rm IR}}
\newcommand{\LmuUV}{L_\mu^{\rm UV}}

\usepackage[dvipsnames]{xcolor}

\begin{document}

\preprint{ZU-TH-22/20}

\title{Vector Leptoquarks Beyond Tree Level II: \texorpdfstring{$\cO(\alpha_s)$}{O(as)} Corrections and Radial Modes}
 
\author{Javier Fuentes-Mart\'{\i}n}
\email{fuentes@physik.uzh.ch}
\affiliation{Physik-Institut, Universit\"at Zu\"rich, CH-8057 Z\"urich, Switzerland}
\author{Gino Isidori}
\email{isidori@physik.uzh.ch}
\affiliation{Physik-Institut, Universit\"at Zu\"rich, CH-8057 Z\"urich, Switzerland}
\author{Matthias K\"{o}nig}
\email{matthias.koenig@uzh.ch}
\affiliation{Physik-Institut, Universit\"at Zu\"rich, CH-8057 Z\"urich, Switzerland}
\author{Nud{\v z}eim Selimovi{\'c}}
\email{nudzeim@physik.uzh.ch}
\affiliation{Physik-Institut, Universit\"at Zu\"rich, CH-8057 Z\"urich, Switzerland}

\begin{abstract}
\vspace{5mm}
We clarify open issues in relating low- and high-energy observables, at next-to-leading order accuracy, in models with a massive leptoquark embedded in a flavor non-universal $SU(4) \times  SU(3) \times SU(2) \times U(1)$ gauge group. Extending previous work on this subject, we present a complete analysis of the  $\cO(\alpha_s)$ corrections to the matching conditions  of semileptonic operators at the high scale. These corrections are not negligible, but they do not exceed the 10\% level and are subleading compared to the $\cO(\alpha_4)$  corrections proportional to the leading leptoquark coupling, which is expected to be much larger than the QCD coupling in the parameter space region of phenomenological interest. We further analyze the impact of radial modes, both at $\cO(\alpha_4)$ and at $\cO(\alpha_s)$ accuracy, highlighting their role in the renormalization of the theory. 
\vspace{3mm}
\end{abstract}

\maketitle

\allowdisplaybreaks


\section{Introduction}\label{sec:intro}

Among the different explanations of the recent $B$-physics anomalies (see e.g.~\cite{deSimone:2020kwi} for a recent review), those based on a Pati-Salam type~\cite{Pati:1974yy} massive vector leptoquark (LQ) turn out to be quite successful from a phenomenological point of view~\cite{Alonso:2015sja,Calibbi:2015kma,Barbieri:2015yvd,Buttazzo:2017ixm,Crivellin:2018yvo}. Various attempts have been made to find a consistent ultraviolet (UV) completion for this field~\cite{Barbieri:2016las,Assad:2017iib,Calibbi:2017qbu,Barbieri:2017tuq,Blanke:2018sro,DiLuzio:2017vat,DiLuzio:2018zxy,Bordone:2017bld,Bordone:2018nbg,Greljo:2018tuh,Cornella:2019hct,Fuentes-Martin:2020bnh, Guadagnoli:2020tlx}. Among them, those based on the gauge group $SU(4)\times SU(3)^\prime\times SU(2)_L \times U(1)_X$~\cite{DiLuzio:2017vat,DiLuzio:2018zxy,Bordone:2017bld,Greljo:2018tuh,Bordone:2018nbg,Cornella:2019hct,Fuentes-Martin:2020bnh,Guadagnoli:2020tlx} (originally proposed in~\cite{Georgi:2016xhm,Diaz:2017lit}, and denoted as ``4321'' in the following) are particularly interesting and well motivated. This is the case especially for those implementations where the SM-like fermions are charged non-universally~\cite{Bordone:2017bld,Greljo:2018tuh,Bordone:2018nbg,Cornella:2019hct,Fuentes-Martin:2020bnh,Guadagnoli:2020tlx}. The interest in such class of models goes beyond their phenomenological impact in $B$-physics: they hint to a possible solution of the Standard Model (SM) flavor puzzle~\cite{Bordone:2017bld}, and might also be able to address the electroweak hierarchy problem~\cite{Fuentes-Martin:2020bnh}.

To investigate the interplay between precision measurements and collider searches in this interesting class of models, it is important to explore the relation between low- and high-energy observables beyond the tree level. In a previous paper~\cite{Fuentes-Martin:2019ign}, we have evaluated the next-to-leading-order (NLO) corrections in the perturbative expansion in the LQ coupling, $\alpha_4 = g_4^2/(4\pi)$, to the matching conditions of semileptonic operators at the high scale. These corrections are sizable, but  still within a perturbative regime even  for $g_4 \approx 3$ (a reference value motivated by the $B$-physics anomalies, see e.g.~\cite{Cornella:2019hct}). The main effect is an enhanced LQ contribution at low-energy, at fixed on-shell coupling, that for $g_4=3$ ranges from $15\%$ to $40\%$, depending on the operator~\cite{Fuentes-Martin:2019ign}. 
 
Our scope here is to analyze two subleading contributions that were not included in~\cite{Fuentes-Martin:2019ign}:
 $\cO(\alpha_s)$ corrections and the contributions from radial modes. Naively, both effects are expected to be well below that of the leading $\mathcal{O}(\alpha_4)$ corrections analyzed in~\cite{Fuentes-Martin:2019ign} (for the phenomenologically motivated large values of the $g_4$ coupling), but a precise assessment of their size is important. 

As far as the $\cO(\alpha_s)$ corrections to LQ-induced semileptonic operators are concerned, a clarification of their size is particularly motivated given the recent analysis in~\cite{Aebischer:2018acj}. Employing an effective vector LQ model that is not UV complete, it is claimed in Ref.~\cite{Aebischer:2018acj} that these corrections exceed the $10\%$ level. As we show, this estimate is far too big for the realistic class of models we are interested in. This is because of two effects:  i)~the presence of additional colored states with mass close to that of the LQ, an unavoidable feature of realistic models addressing the $B$-physics anomalies~\cite{DiLuzio:2017vat,Baker:2019sli}; ii)~the lack of a physical renormalization condition for the LQ coupling. A  proper treatment of both effects significantly reduce the impact of the QCD corrections.

As far as radial modes are concerned, a clarification of their role in the renormalization of the theory provides a motivation for a deeper investigation of their effects.  As we show in this paper, one can completely decouple these states only if the spectrum of the heavy vectors is degenerate (in the so-called $SU(4)_V$ custodial limit). If this is not the case, the infinite mass limit for the radial modes leads to ambiguities in the estimate of NLO effects that can be cured at the price of introducing independent renormalization conditions for the couplings of the different heavy vectors.  However, we demonstrate that for realistic spectra, with large but not infinite masses for the radials, the size of these ambiguities is negligible already at $\cO(\alpha_4)$.

\section{The model}\label{sec:model}

\begin{table}[t]
\begin{center}
\setlength{\tabcolsep}{5pt}
\renewcommand{\arraystretch}{1.1}
\begin{tabular}{|c|c|c|c|c|c|c|}
\hline
Field & $SU(4)$ & $SU(3)^\prime$ &  $SU(2)_L$  & $U(1)_X$ \\
\hline
\hline 
 $\psi_L$ & $\mathbf{4}$  & $\mathbf{1}$  & $\mathbf{2}$  &  0 \\
$\psi_R^+$ & $\mathbf{4}$ &  $\mathbf{1}$   & $\mathbf{1}$  &  $ 1/2$  \\[3pt] 
$\psi_R^-$ & $\mathbf{4}$ &  $\mathbf{1}$   & $\mathbf{1}$  &  $-1/2$  \\[3pt] 
$q_L^i$ & $\mathbf{1}$ &  $\mathbf{3}$   & $\mathbf{2}$  &  $1/6$   \\[3pt] 
$u_R^i$ & $\mathbf{1}$ &  $\mathbf{3}$   & $\mathbf{1}$  &  $2/3$  \\[3pt] 
$d_R^i$ & $\mathbf{1}$ &  $\mathbf{3}$   & $\mathbf{1}$  &  $-1/3$  \\[3pt]  
\hline
\hline
$\Omega_3$ & $\mathbf{\bar 4}$ &  $\mathbf{3}$   & $\mathbf{1}$  &  $1/6$  \\
$\Omega_1$ & $\mathbf{\bar 4}$ &  $\mathbf{1}$   & $\mathbf{1}$  &  $-1/2$  \\  
\hline
\end{tabular}
\end{center}
\caption{\sf Matter fields with  non trivial charges under $SU(4) \times SU(3)^\prime$. Here $i=1,2$, $\psi_L\equiv(q_L^3\; \ell_L^3)^\intercal$, $\psi_R^+\equiv(u_R^3\; \nu_R^ 3)^\intercal$ and $\psi_R^-\equiv(d_R^3\; e_R^3)^\intercal$.}
\label{tab:content}
\end{table}

We consider a model based on the $SU(4)\times SU(3)^\prime\times SU(2)_L\times U(1)_X$ gauge group. This symmetry contains the SM gauge group as a subgroup: QCD corresponds to the vectorial $SU(3)$ subgroup of $SU(4)\times SU(3)^\prime$, $SU(2)_L$ is as in the SM, and the hypercharge is defined in terms of the $U(1)_X$ charge $X$, and the $SU(4)$ generator $T^{15}=1/(2\sqrt{6})\,\mathrm{diag}(1,1,1,-3)$ by $Y=X+\sqrt{2/3}\,T^{15}$. Since in this paper we are interested in NLO corrections in the $SU(4)$ and $SU(3)^\prime$ couplings, we simplify the analysis by setting the $SU(2)_L\times U(1)_X$ gauge couplings to zero and we do not consider the SM Higgs sector. We denote the $SU(4)\times SU(3)^\prime$ gauge couplings by $g_4$ and $g_3$, respectively, and the corresponding gauge fields by $H_\mu^A$ and $C_\mu^a$, with $A=1,\dots,15$ and $a=1,\dots,8$. The SM gluons, $G^a_\mu$, and the QCD coupling, $g_s$, are given in terms of the $SU(4) \times SU(3)^\prime$ couplings and fields by
\begin{align}
G^a_\mu= \left( s_3\, H^a_\mu + c_3\,C^a_\mu\right), \quad  
g_s = g_3\, c_3 = g_4\, s_3,
\end{align}
where $c_3 \equiv  \cos\theta_3= g_4/\sqrt{ g_4^2 +g_3^2}$, and $s_3 \equiv  \sin\theta_3= g_3/\sqrt{ g_4^2 +g_3^2}$, and therefore $c_3 = \sqrt{1- g_s^2/g_4^2}$. The additional gauge bosons transform under the SM subgroup as $U\sim(\boldsymbol{3},\boldsymbol{1},2/3)$, $G^\prime\sim(\boldsymbol{8},\boldsymbol{1},0)$ and $Z^\prime\sim(\boldsymbol{1},\boldsymbol{1},0)$. In terms of the original gauge bosons, they are given by
\begin{align}
G^{\prime\,a}_{\mu}&= c_3\, H^a_\mu- s_3\, C^a_\mu  \,,\qquad  Z^\prime_\mu=H_\mu^{15}\,,  \no\\
U_\mu^{1,2,3}&=\frac{1}{\sqrt{2}}\left(H_\mu^{9,11,13}-iH_\mu^{10,12,14}\right)\,.
\end{align}
These become massive after the spontaneous symmetry breaking of $SU(4)\times SU(3)^\prime\to SU(3)_c$. The symmetry breaking is triggered by the vacuum expectation value (vev) of two scalar fields transforming in the anti-fundamental of $SU(4)$, $\Omega_1$ and  $\Omega_3$, singlet and triplet under $SU(3)^\prime$, respectively. The gauge boson masses read
\begin{align}
m_G&=\frac{g_4}{2c_3}\,\sqrt{2\omega^2_3}\,,  \qquad 
\mZ=\frac{g_4}{2}  \sqrt{\frac{3\omega_1^2+  \omega_3^2}{2}}\,,\no\\
m_U&= \frac{g_4}{2} \sqrt{\omega_1^2+\omega_3^2}\,,
\end{align}
with $\omega_{1,3}$ denoting the $\Omega_{1,3}$ vevs. In the limit $\omega_1=\omega_3$ and $g_3=0$, there is a residual custodial $SU(4)_V$ global symmetry.\footnote{We denote the custodial symmetry $SU(4)_V$ by the vectorial subgroup of $SU(4)\times SU(4)^\prime$, with $SU(4)^\prime \supset  SU(3)^\prime$. In the custodially preserving limit, $\Omega_1$ and $\Omega_3$ can be seen as two components of a single field $\Omega_4$, transforming as $4\times \bar 4$ under $SU(4)\times SU(4)^\prime$ and responsible for the breaking $SU(4)\times SU(4)^\prime \to SU(4)_V$.}

The $\Omega_{1,3}$ fields decompose under the SM subgroup as 
\begin{widetext}
\begin{align}\label{eq:Omega13}
\Omega_1^\dagger&=\frac{\omega_1}{\sqrt{2}}
\begin{pmatrix}
\frac{g_4}{\sqrt{2}}\frac{1}{m_U}\big(\phi_U+\cot\beta\, h_U\big) \\[10pt]
1+\frac{S_1}{\omega_1}+i\,\frac{-3g_4}{2\sqrt{6}}\frac{1}{m_{Z^\prime}}\big(\phi_{Z^\prime}-\frac{1}{\sqrt{3}}\cot\beta\,h_{Z^\prime}\big)
\end{pmatrix}
\,,\no\\[5pt]
\Omega_3^\dagger&=\frac{\omega_3}{\sqrt{2}}
\begin{pmatrix}
\mathbb{1}_{3\times3}\,\big[1+\frac{1}{\sqrt{3}}\frac{S_3}{\omega_3}+i\,\frac{g_4}{2\sqrt{6}}\frac{1}{m_{Z^\prime}}\big(\phi_{Z^\prime}+\sqrt{3}\tan\beta\,h_{Z^\prime}\big)\big]+T^a\,\frac{\sqrt{2}}{\omega_3}\,\big(i\,\phi^a_{G^\prime}+h_{G^\prime}^a\big) \\[10pt]
-\frac{g_4}{\sqrt{2}}\frac{1}{m_U}\big(\phi_U^\dagger-\tan\beta\,h_U^\dagger\big) 
\end{pmatrix}
\,,
\end{align}
\end{widetext}
with $\tan\beta=\omega_1/\omega_3$, and where $\phi_i$ and $h_i$ are, respectively, would-be Goldstone and radial modes with the same quantum numbers as the corresponding gauge fields, and $S_1$ and $S_3$ are SM-singlet radial modes. An explicit expression for the renormalizable potential of $\Omega_{1,3}$ can be found in~\cite{DiLuzio:2018zxy}. An interesting limit, realized in explicit frameworks such as the composite model in~\cite{Fuentes-Martin:2020bnh}, is that of heavy radial modes  (i.e.~$M^2_R \gg g^2_4\, \omega^2_{3(1)})$. In this limit, the effect of these fields decouple and we are left with a non-linear realization of the $SU(4) \times SU(3)^\prime\to SU(3)_c$  symmetry breaking. As we show in Section~\ref{sec:radials}, an ample class of physical observables can be evaluated to NLO accuracy with no (or marginal) ambiguities also in the non-linear case. 

We focus on an implementation where the fermion sector is arranged in the following flavor non-universal way: one generation is $SU(3)^\prime$ singlet and transforms as Pati-Salam representations under $SU(4)\times SU(2)_L\times U(1)_X$, while the other two generations are $SU(4)$ singlets and have SM charges under $SU(3)^\prime\times SU(2)_L \times U(1)_X$. In the absence of fermion mixing effects, which we do not consider here, the fermions charged under $SU(4)$ are identified with the SM third family (plus a right-handed neutrino). We note that, although we focus on this specific implementation, our results can be applied with minimal changes to other models considered in the literature, like the one in~\cite{DiLuzio:2017vat,DiLuzio:2018zxy}. The matter content (fermions and scalars) with non trivial transformation properties under $SU(4)\times SU(3)^\prime$ is summarized in Table~\ref{tab:content}. 

Finally, the interactions between the heavy vectors and fermions read
\bea
&& \cL_{\rm int} \supset    \frac{g_4}{\sqrt{2}}\, U_\mu\,\bar\psi_q\,\gamma^\mu\psi_\ell  +{\rm h.c.} \no\\
&& \qquad + \frac{g_4}{2\sqrt{6}}\, Z^\prime_\mu \,(\bar \psi_q\gamma^\mu\, \psi_q-3\,\bar \psi_\ell\gamma^\mu\, \psi_\ell)    \\
&& \qquad + g_4\, c_3\, G^{\prime\,a}_\mu  \bigg[  \bar\psi_q\gamma^\mu\,T^a\psi_q  
 - \frac{s_3^2}{c_3^2}   \sum_{i=1,2}  \bar q^{\prime\, i } \gamma^\mu\,T^a   q^{\prime\, i } \bigg],  \no
\label{eq:U1Lag}
\eea
where  
$\psi = (\psi_q\;\psi_\ell)^\intercal$ generically denote the $SU(4)$ fermion multiplets, $q= (u\;d)^\intercal$ contains both chiralities, and $T^a$ are the $SU(3)$ generators.

\section{Matching corrections at \texorpdfstring{$\cO(\alpha_s)$}{O(as)}}\label{sec:nlo}

\subsection{General strategy and operator basis}
In this section, we evaluate the $\alpha_s(m^2_U)$ corrections to the Wilson coefficients of the dimension-six semileptonic operators involving SM third-generation fermions, and in the limit of heavy radial modes. Following Ref.~\cite{Fuentes-Martin:2019ign}, we adopt the following normalization
\begin{align}
 \mathcal L = -\frac{g_4^2}{2m_U^2}\sum_k \mathcal C_k(\mu) \mathcal O_k\,,
\end{align}
and operator basis
\bea
 \cO_{LL}^U &=& (\bar \ell_L^3 \gamma^\mu q_L^3)(\bar q_L^3 \gamma_\mu \ell_L^3)\,, \no\\ 
 \cO_{LR}^{U} &=& -2(\bar \ell_L^3 e_R^3)(\bar d_R^3 q_L^3) + \mathrm{h.c.}\,,  \no \\
 \cO_{ld} &=& (\bar \ell_L^3\gamma^\mu \ell_L^3)(\bar d_R^3 \gamma_\mu d_R^3)\,, \no\\
 \cO^{(1)}_{lq} &=& (\bar \ell_L^3 \gamma^\mu \ell_L^3)(\bar q_L^3\gamma_\mu q_L^3 )\,,\no \\ 
 \cO_{qe} &=& (\bar q_L^3 \gamma^\mu q_L^3)(\bar e_R^3 \gamma_\mu e_R^3)\,,  \no\\
 \cO_{de} &=& (\bar d_L^3 \gamma^\mu d_L^3)(\bar e_R^3 \gamma_\mu e_R^3)\,,  
 \label{eq:EFTops}
 \eea
where the quark and lepton fields belong to the $SU(4)$ charged fermions in Table~\ref{tab:content}. The tree-level expression for the Wilson coefficients read
\begin{align}
\cC^{U}_{LL} &=  \cC^{U}_{LR} = 1\,,   & [ U_1~~{\rm exchange}]\no \\
\cC_{lq}^{(1)} &= \cC_{ld} =  \cC_{qe}  = -\frac{1}{4x_{Z^\prime}}\,,   &[ Z^\prime~{\rm exchange}]\no\\
\cC_{de} &= 1 -\frac{1}{4x_{Z^\prime}}\,,   &[ U_1~\&~Z^\prime~{\rm exchange}]
\label{eq:Ctree}
\end{align}
where $x_{Z^\prime} = m_{Z^\prime}^2/m_U^2$.

The NLO corrections to the semileptonic amplitudes we are interested in can be divided into two main categories: factorizable contributions (due to vertex and self-energy diagrams), which enter as multiplicative corrections to a given tree-level amplitude, and non-factorizable corrections, arising from box diagrams. In Ref.~\cite{Fuentes-Martin:2019ign}, we evaluated the $\alpha_4$ corrections employing an on-shell renormalization scheme where, by construction, the factorizable corrections vanish on-shell. In the case of the $\alpha_s$ corrections, it is more convenient to perform the calculation in the \msbar\ scheme, both because of infrared (IR) singularities and because of the precise knowledge of $\alpha_s$ in the \msbar\ scheme (at arbitrary scales). However, proceeding this way, we need to take an additional correction factor into account in order to express the (unphysical) coupling $g_4$ appearing in the (high-scale) Lagrangian in terms of some high-energy physical observable: we choose the latter to be the inclusive LQ decay width, and we denote the corresponding physical coupling as $\hat g_4$.

The calculation is performed in the Feynman gauge, treating $g_4$ and $g_s$ as two independent couplings, and retaining only corrections of $\cO(\alpha_s)$, given the $\cO(\alpha_4)$ contributions have already been derived in~\cite{Fuentes-Martin:2019ign}.  
The amplitudes in the full theory receive contributions from both gluon- and coloron-exchange diagrams (as well as the corresponding ghost and Goldstone fields, in the case of the vector self energy). For each gluon-mediated amplitude, there is a corresponding coloron-mediated amplitude whose coupling is proportional to $(g_4  c_3)^2 = g_4^2 -g_s^2$. This ensures a perfect cancellation of the UV divergences proportional to $g_s^2$. Such result is expected since we are interested in amplitudes involving external states which are charged only under $SU(4)$. More precisely, the structure of the $\cO(\alpha_s)$ amplitudes we are interested in, summing coloron- and gluon-mediated contributions, is of the type
\be
\left. \cA (s) \right|_{\cO(g_4 g_s^2)} = g_s^2 \left [  \cA^{(g)} (s) -  \cA^{(\Gp)} (s) \right] \,,
\ee
such that the result is finite and vanishes identically in the limit $v_3 \to0$.\footnote{In the limit $v_3=0$ the gauge symmetry is spontaneously broken into  $SU(3)_4 \times SU(3)^\prime$, and the $SU(3)^\prime$ group is irrelevant for the amplitudes we are looking at.}

Unless otherwise specified, the results presented in  this section are reported in the limit $m_{\Gp}=m_U$.\footnote{For $g_3\not=0$, the limit  $m_{\Gp}=m_U$ does not correspond to the $SU(4)_V$ custodial limit. However, this does not cause a problem in the evaluation of these amplitudes, which are finite independently of the radial modes. The limit $m_{\Gp}=m_U$ is then particularly convenient in order to obtain simple expressions.} Complete expressions for generic masses and generic momentum transfer for the LQ two-point and vertex corrections can be found in Appendix~\ref{sect:app}.

\subsection{Renormalization of \texorpdfstring{$g_4$}{g4} from the LQ width}
The amplitude for $U\to q\ell$, at NLO accuracy in $\alpha_s$, can be decomposed as
	\bea
	\cA^U_{\rm NLO} & =& \cA_{\rm tree}
	\Big[1+\frac{\alpha_s}{4\pi}\Big(\delta V_U(\us)     \no \\
&&	 \qquad + \frac{1}{2}\delta Z_{U}(\us) + \frac{1}{2}\delta Z_{q}(0)\Big)\Big]~.
    \eea
Summing the contributions from gluon- and coloron-exchange, the quark wave-function correction is 
\bea
 \delta Z_{q}(0) &=&  C_F\left(\frac{1}{\epsIR}+ \LmuIR -\log x_{\Gp} - \frac{1}{2}  \right)~,
 \label{eq:Sigmaq}
\eea
where $\LmuIR=\log(\mu_{\rm IR}^2/m_U^2)$ and, for illustrative purposes, we have kept the explicit dependence on $m_{\Gp}$. As anticipated, the result is UV finite. The UV divergence occurring in pure QCD is recovered in the limit of infinite coloron mass.

As far as the LQ two-point function is concerned, we renormalize the mass on-shell
(i.e.~we directly express $m_U$ in terms of the physical LQ mass). Decomposing the one-loop contribution to the two-point function as 
\be
i \Sigma_U^{\mu\nu}(q^2) = i g^{\mu\nu} \frac{\alpha_s}{4\pi} \Sigma_U (q^2) + \cO(q^\mu q^\nu)~,
\ee
we define the wave-function correction as
\be
\delta  Z_{U}(s) = \frac{\Sigma_{U}(s) - \Sigma_{U}(m_U^2) }{s-\us}.
\ee
For $s=\us$, setting $m_{\Gp}=m_U$, we find
\be 
 \lim_{s\to\us}  \delta  Z_{U}(s)
 = - \frac{8}{3}\left(\frac{1}{\epsIR}+ \LmuIR \right)  
 + \frac{56}{9}- \frac{8}{9} \sqrt{3}\pi~. \quad 
\ee
Finally, the vertex correction, computed also in the limit  $m_{\Gp}=m_U$, is 
 \bea
	\delta V_U(\us) &=& C_F \Bigg[-2 \left(\frac{1}{\epsIR}  + \LmuIR \right)  - \frac{1}{\epsIR} \LmuIR - \frac{1}{\epsIR^2}
	\no\\
&&	-\frac{1}{2}(\LmuIR)^2 -1-\sqrt{3}\pi+\frac{7}{12}\pi^2\Bigg]~.
\eea

Using the above results, the two-body contribution to the decay amplitude, including NLO virtual corrections in a $d$-dimensional phase space, can be expressed as 
\bea
&& \Gamma^V  = \frac{g_4^2}{2 m_U}\frac{1}{(d-1)} \left(\frac{e^{\gamma_E} \mu^2}{m_U^2} \right)^{\frac{4-d}{2}}\frac{2^{-d}(d-2)}{\sqrt{\pi}\Gamma(\frac{d-1}{2})}   \no \\
&& \quad \times \Big[ 1+\frac{\alpha_s}{4\pi}\left(2\delta V_U(\us)+ \delta Z_{U}(\us) + \delta Z_{q}(0)\right)\Big]. \quad 
\eea
The corresponding contribution due to real radiation, integrated over the whole phase space is 
\bea
\Gamma^R  &=& \frac{g_4^2}{4\pi}\,\frac{m_U}{(d-1)}\,\frac{\alpha_s }{4\pi}C_F  \Big(\frac{1}{\epsIR^2} 
	+ \frac{7}{2 \epsIR } +\frac{2}{\epsIR} \LmuIR    \no \\
	&& + 7  \LmuIR + 2 (\LmuIR)^2 -\frac{5\pi^2}{6} + \frac{155}{12}\Big)~.
\eea
Summing these two contributions, we obtain the (IR-finite) expression for the NLO inclusive decay width
\bea
\Gamma_{\rm incl} &=& \frac{g_4^2 m_U }{24\pi}\left[ 1
	+\frac{\alpha_s}{4\pi}\Bigg( \frac{76}{3}-  \frac{32}{9}  \sqrt{3}\pi  \Bigg)\right]
	\no\\
	&\equiv&  \frac{\hat g_4^2 m_U }{24\pi}~,
\eea
from which we define relation between the physical coupling $\hat g_4$ and $g_4$:
\be
	g_4 = \hat g_4  \left( 1+   \frac{\alpha_s}{4\pi} \delta_g \right),  \quad 
	\delta_g = -  \frac{38}{3} +  \frac{16}{9}  \sqrt{3}\pi~.
	\label{eq:deltag}
\ee

\subsection{Four-fermion amplitudes and Wilson coefficients}
The factorizable NLO corrections to the LQ exchange amplitude in the zero momentum limit can be written as
\bea
\mathcal{A}^{U}_{\text{NLO}} &=& 
\mathcal{A}^{U}_{\text{tree}} \left( 1+\frac{\alpha_s}{4\pi} \kappa_U \right)~, \no\\
 \kappa_U &=&
2\delta_g + 2 \delta V_U(0) + \delta  Z_U(0) +  \delta Z_{q}(0)~.
\eea
The coupling renormalization and quark wave-function terms are given in~\eqref{eq:deltag} and~\eqref{eq:Sigmaq}, respectively, while the LQ wave-function and vertex at zero momentum transfer are
\bea
\delta  Z_U (0) &=& \frac{ \Sigma_{U}(m^2) - \Sigma_{U}(0) }{\us} = -\frac{100}{9}+\frac{32}{9} \sqrt{3} \pi\,, \quad  \no\\
\delta V_U(0) &=& 4\,.
\eea
Putting all pieces together, we arrive at
\be 
 \kappa_U  =  C_F \left(\frac{1}{\epsIR}+\LmuIR \right)
 -\frac{262}{9} + \frac{64}{9} \sqrt{3} \pi~. 
\ee

Proceeding in analogous fashion, we compute the factorizable corrections to the $Z^\prime$ exchange amplitude. Here, there is no $\cO(\alpha_s)$ correction to the vector two point function. Moreover, the quark vertex correction at zero momentum transfer satisfies
\be
\delta V^{q\bar q}_{\Zp}(0) = - \delta Z_{q}(0)~,
\ee
as expected from the conservation of the vector current in QCD. We thus have
\be
 \kappa_{\Zp} =   2\delta_g + \delta V^{q\bar q}_{\Zp}(0)  + \delta Z_{q}(0) = 
   2\delta_g~,
 \ee
and therefore the only effect is the change in the overall coupling once we express it in terms of $\hat g_4$.

The last necessary ingredient to evaluate the amplitudes in the full theory are the box diagrams. We present these amplitudes projecting the result into the EFT basis,
\be
\cA^{\rm box} =  -\frac{g_4^2}{2m_U^2}\sum_k   \frac{\alpha_s}{4\pi}\,  \delta C^{\rm box}_k 
\langle \mathcal O_k \rangle~.
\ee
Separating the non-factorizable $\delta C^{\rm box}_k$ contributions into leptoquark-gluon and leptoquark-coloron amplitudes,
\be
 \delta C^{\rm box}_k = \delta C^{[U g]}_k+ \delta C^{[U \Gp]}_k~,
\ee
we find
\bea
&& \delta C^{[U \Gp]}_{LR} = 4 \delta C^{[U \Gp]}_{LL} = 4 \delta C^{[U \Gp]}_{de} =  - 4
 C_F  f(\xG)~,   \no \\
&& \delta C^{[U g]}_{LL} = \delta C^{[U g]}_{de} = 
-   C_F  \left(\frac{1}{\epsIR}+\LmuIR  - \frac{1}{2} \right)~, \no \\
&& \delta C^{[U g]}_{LR} = 
-   4C_F \left(\frac{1}{\epsIR}+\LmuIR  +  \frac{1}{2} \right)~,   \no \\
&& \delta C^{\rm box}_{ld} =  \delta C^{(1), {\rm box}}_{lq} =  \delta C^{\rm box}_{qe}  = 0~,
\eea
where $f(x)= \log x/(x-1)$ and $f(1)=1$.

To obtain the $\cO(\alpha_s)$ corrections to the Wilson coefficients, we need to perform the matching between the full and the effective theory. This is trivial for all the operators in  (\ref{eq:EFTops}) except for $\cO_{LR}$, since they involve only conserved currents (in QCD). In the case of $\cO_{LR}$, we need to subtract the $\cO(\alpha_s)$ corrections to the matrix element in the EFT from the full amplitude. This yields the additional term
\be
 \delta C^{\rm match}_{LR} =  3 C_F \left( \frac{1}{\epsIR} + \LmuIR - L_\mu \right)~,
\ee
where the scale $\mu$ appearing in $L_\mu$ is the matching scale. 

Putting all the ingredients together, we can write the $\cO(\alpha_s)$ corrections to the Wilson coefficients  as 
\bea
\cC^{U}_{LL} &=& 1+ \frac{\alpha_s}{4\pi}\left(  \kappa_U  + \delta C^{\rm box}_{LL} \right)\,,  \no \\
\cC^{U}_{LR} &=& 1  + \frac{\alpha_s}{4\pi}\left(  \kappa_U  + \delta C^{\rm box}_{LR} + 
 \delta C^{\rm match}_{LR} (\mu) \right)\,, \no\\
\cC_{de} &=& 1 + \frac{\alpha_s}{4\pi}\left(  \kappa_U  + \delta C^{\rm box}_{de} \right)
 -\frac{1}{4x_{Z^\prime}} \left( 1+ \frac{\alpha_s}{4\pi}  \kappa_\Zp\right)~, \no \\
 \cC_{ld} &=&  C_{lq}^{(1)}=  \cC_{qe}  = -\frac{1}{4x_{Z^\prime}} \left( 1+ \frac{\alpha_s}{4\pi}  \kappa_\Zp \right)~.
\eea
In the $SU(4)_V$ custodial limit, this leads to 
\bea
\cC^U_{LL} &=& 1+ \frac{\alpha_s}{ 4\pi}\left(  -\frac{268}{9}+ \frac{64}{9}\sqrt{3}\pi   \right)\,, \no \\
\cC^U_{LR} &=&  1 + \frac{\alpha_s}{4 \pi}\left( -4 L_{\mu } -\frac{334}{9} + \frac{64}{9}\sqrt{3}\pi  \right) \,,  \no\\
\cC_{de} &=& \frac{3}{4}+ \frac{\alpha_s}{ 4\pi}\left(
-\frac{211}{9} + \frac{56}{9}   \sqrt{3} \pi \right)\,, \no \\ 
 \cC_{ld} &=&  C_{lq}^{(1)}=  \cC_{qe}  =  - \frac{1}{4}+ \frac{\alpha_s}{ 4\pi}\left(
   \frac{19}{3}- \frac{8}{9} \sqrt{3} \pi \right). \quad 
\eea
As expected, the scale dependence of $\cC^U_{LR}$ matches the one expected from the anomalous dimension of this operator at $\cO(\alpha_s)$.

\subsection{Discussion}

Including also the $\cO(\alpha_4)$ terms computed in~\cite{Fuentes-Martin:2019ign},
the NLO expressions of the two Wilson coefficients containing charged currents are
\bea
	\cC^U_{LL}  &=& -\frac{\gp^2}{2\us}\left( 1+\frac{ \alpha_4}{4\pi} c^{(4)}_{LL}  	
	+\frac{ \alpha_s}{4\pi} c^{(s)}_{LL}   \right), \no\\
	\cC^U_{LR}  &=& -\frac{\gp^2}{2\us}\left[ 1+ \frac{\alpha_4}{4\pi} c^{(4)}_{LR}    
	+\frac{ \alpha_s}{4\pi} \left( -4 L_\mu +  c^{(s)}_{LR}  \right)  \right], \quad
\eea
where 
\bea
&& c^{(4)}_{LL}  = \frac{829}{24}-11\sqrt{3}\pi+\frac{17\pi^2}{6}  \approx +2.65~, \no\\
&& c^{(s)}_{LL}  =   -\frac{268}{9}+ \frac{64}{9}\sqrt{3}\pi  \approx +8.92~, \no\\
&&  c^{(4)}_{LR}  = \frac{937}{24}-11\sqrt{3}\pi+\frac{17\pi^2}{6} \approx +7.15~,  \no\\
&&  c^{(s)}_{LR}  = -\frac{334}{9} + \frac{64}{9}\sqrt{3}\pi  \approx +1.58~.  
\eea
Two main observations are in order:
\begin{itemize}
\item[i)] All matching corrections are positive. Hence, the enhancement of low-energy amplitudes at fixed high-energy inputs already noted in~\cite{Fuentes-Martin:2019ign} is further strengthened by the inclusion of the $\cO(\alpha_s)$ terms. 
\item[ii)] The coefficients of the $\cO(\alpha_4)$ and $\cO(\alpha_s)$ corrections are of similar size. Since we expect $\alpha_4 \gg \alpha_s$, this implies that the $\cO(\alpha_s)$ corrections are clearly subdominant. The largest $\cO(\alpha_s)$ correction is the one to $\cC^U_{LL}$, which does not exceed $6\%$ for  $\alpha_s(2~{\rm TeV})=0.083$.
\end{itemize} 

The second statement apparently contradicts the result of Ref.~\cite{Aebischer:2018acj}, where it is claimed that QCD matching corrections in $\cC^U_{LL}$ exceed the 10\% level. However, this difference can be understood from the fact that the calculation made in Ref.~\cite{Aebischer:2018acj} differs from our analysis in two important points.
First, the vector LQ in Ref.~\cite{Aebischer:2018acj} is introduced using an effective approach, without embedding it into a complete UV theory. This is not consistent for this field. In the realistic case we have discussed, the embedding into the 4321 model brings the additional contribution of the coloron, providing a consistent cut-off for all the apparent UV singularities appearing in the effective approach. Second, the effective LQ coupling in Ref.~\cite{Aebischer:2018acj} is treated as a generic \msbar\ coupling and no attempt is made to connect it to physical observables. Proceeding this way, the analysis of Ref.~\cite{Aebischer:2018acj} does not take into account the sizable corrections due to the relation between $g_4$ and high-energy observables.

We have explicitly checked that our results are consistent with those of Ref.~\cite{Aebischer:2018acj} once these two effects are taken into account. We can indeed decompose the finite terms $c^{(s)}_{LL,LR}$ as follows:
\be
c^{(s)}_{LL,LR} =  \bar c^{(s,g)}_{LL,LR} +  \bar c^{(s,\Gp)}_{LL,LR} + 2 \left( \delta_g^{(s,g)} + \delta_g^{(s,\Gp)} \right)\,,
\ee
where $\bar c^{(s,g)}_{LL,LR}$ are the  \msbar\ results (still with on-shell renormalization for the LQ mass), obtained in the absence of coloron contributions and coupling renormalization. These agree with the terms reported in Ref.~\cite{Aebischer:2018acj}. Subtracting the (unphysical) scale dependence related to the \msbar\ renormalization procedure, they are given by
\begin{align}
 \bar c^{(s,g)}_{LL} -  \frac{100}{9} L_\mu =  \frac{482}{27}  \approx  +17.9~, \no \\
 \bar c^{(s,g)}_{LR} - \frac{100}{9} L_\mu =  \frac{392}{27}  \approx  +14.5~. 
 \end{align}
These apparently large corrections are significantly reduced  by the coloron contributions and by the coupling renormalization. We further split the latter into gluon-induced and coloron-induced terms:
\bea
 \bar c^{(s,\Gp)}_{LL}  +\frac{100}{9} L_\mu &=&  - \frac{602}{27} + \frac{32}{9} \sqrt{3}\pi \approx - 2.95  \,, \no\\
 \bar c^{(s,\Gp)}_{LR} + \frac{100}{9} L_\mu  &=&  -\frac{710}{27} + \frac{32}{9} \sqrt{3}\pi \approx  - 6.95   \, \no\\ 
 \delta_g^{(s,\Gp)} - \frac{50}{9} L_\mu   &=& -\frac{11}{27}  + \frac{16}{9} \sqrt{3}\pi -  \frac{8}{9}\pi^2    \approx  + 0.49~, \no\\
  \delta_g^{(s,g)} + \frac{50}{9} L_\mu   &=& -\frac{331}{27} + \frac{8}{9}\pi^2 \approx  - 3.49~.
\eea
As can be seen, in the case of  $c^{(s)}_{LR}$ the effect of the coloron and of the coupling renormalization are very similar in size, and imply a reduction of the gluon-only \msbar\ result of almost one order of magnitude. The reduction is less pronounced in the case of $c^{(s)}_{LL}$, but still results in an almost $50\%$ reduction compared to Ref.~\cite{Aebischer:2018acj}.

\section{Impact of radial modes}\label{sec:radials}

The purpose of this section is to elucidate the role of radial modes, with particular reference to their heavy-mass limit. In the absence of Yukawa couplings, the radial modes enter the amplitudes we are interested in only via the two-point functions of the heavy vectors. Focusing the attention on the $\cO(\alpha_4)$ corrections, we define 
\be
i \Sigma_V^{\mu\nu}(q^2) = i g^{\mu\nu} \left[ \frac{\alpha_4}{4\pi} \Sigma^{(4)}_V (q^2) + \cO(\alpha_s) \right]  +
\cO(q^\mu q^\nu)~,
\ee
and we decompose $\Sigma^{(4)}_V (s)$  as 
\be
\Sigma^{(4)}_V (s) = \Sigma^{(4)}_{V,0} + s\, \Sigma^{(4)}_{V,1} +  \Sigma^{(4)}_{V,{\rm reg}} (s)~, 
\label{eq:twopoint4}
\ee
where $\Sigma^{(4)}_{V,{\rm reg}} (s)$ is regular at $s=0$ and is free from UV 
divergences.\footnote{Complete expressions for  $\Sigma^{(4)}_{V,1}$ and  $\Sigma^{(4)}_{V,{\rm reg}} (s)$, in the absence of radial fields,  are reported in Appendix~\ref{sect:app}. These expressions complement that of $\Sigma^{(4)}_{U,{\rm reg}} (s)$ already presented in \cite{Fuentes-Martin:2019ign}.}
 
In a renormalizable theory, as it is the case for our model once we include the radial modes, the  power-like and logarithmic non-decoupling contributions from heavy degrees of freedom can always be absorbed into a redefinition of $\Sigma^{(4)}_{V,0}$ and $\Sigma^{(4)}_{V,1}$. Employing an on-shell renormalization scheme for the vector masses, we can ignore this effect in $\Sigma^{(4)}_{V,0}$ by absorbing it into the physical masses. Therefore, the only terms where radial modes can exhibit a non-decoupling behavior in the heavy mass limit are the wave-function corrections, $\Sigma^{(4)}_{V,1}$.

If we are interested in processes involving a single heavy vector, this sensitivity to $\Sigma^{(4)}_{V,1}$ is eliminated by the on-shell renormalization of the corresponding heavy vector coupling, as has been done in \cite{Fuentes-Martin:2019ign} for the LQ case. However, a potential problem arises if we are interested in processes where two or more heavy vectors are involved already at the tree-level, since two or more $\Sigma^{(4)}_{V,1}$
appear at NLO.

The UV divergent terms in $\Sigma^{(4)}_{V,1}$ have the following structure 
\be
\Sigma^{(4)}_{V,1} =  \left( z^{\rm YM}  +  z^{\phi}_{V} +  z^{R}_{V}  \right) \left( \frac{1}{\epsUV} +\LmuUV \right)~,
\ee
with $V=U,Z^\prime, G^\prime$, and where the three expressions in parentheses refer to the contributions of gauge, fermion, and ghost fields ($z^{\rm YM}$), the contribution of the Goldstone bosons ($z_V^{\phi}$), and the contribution of the radial modes ($z^{R}_{V}$). The gauge, fermion, and ghost factor $z^{\rm YM}=16/3$ is universal. In general, this is not the case for the two separate scalar factors. The Goldstone terms, which are the only contributions included in the expressions reported in the Appendix, are 
\bea
z^{\phi}_{U} &=& - \frac{1}{12} \frac{ \omega^2_1+3 \omega^2_3}{\omega_1^2+\omega_3^2 }~,  \no \\
z^{\phi}_{\Zp} &=& -  \frac{1}{24} \frac{ (\omega^2_1+3 \omega^2_3)^2}{(\omega_1^2+\omega_3^2)^2 }~,  \no \\
z^{\phi}_{\Gp} &=& - \frac{1}{24} \frac{ 7 \omega^4_1+6 \omega^2_1 \omega_3^2 +3\omega_3^2 }{(\omega_1^2+\omega_3^2)^2 }~.
\eea
This implies that, in general, in the non-linear description of the $SU(4) \times SU(3)^\prime \to SU(3)_c$ symmetry breaking, we need to renormalize separately the different vector fields to cancel all UV divergences. This unavoidably decrease the predictive power of the theory, since we need to treat the three couplings as independent free parameters. 

This problem disappears once we include the radial modes. Indeed, the $z^{R}_{V}$ can be written as
\be
z^{R}_{V}  = - \frac{2}{3}-   z^{\phi}_{V}~,
\ee
so their presence restores the universality of the UV divergences, as expected by the renormalizability of the theory. However, this occurs at the cost of introducing (non-decoupling) $\log(M^2_R/m^2_V)$ contributions after the (universal) field renormalization. We have checked explicitly that these non-decoupling effects are in one-to-one correspondence with the residual scale dependence induced by the non-universal behavior of the $z^{\phi}_{V}$ piece. Namely, in the limit of a heavy and degenerate radial spectrum, the non-decoupling $\log(M^2_R/m^2_V)$ effects surviving after renormalization are proportional to the differences among the $z^{\phi}_{V}$.

A key observation is that an effective universal field renormalization is possible also 
in absence of radial fields if $\omega_1=\omega_3$. Indeed, in the $SU(4)_V$ custodial limit, we have
\be
\lim_{\omega_1=\omega_3}  z^{\phi}_{V} = - \frac{1}{6}~.
\ee
In the non-linear description, the different field renormalization for generic $\omega_i$ is a consequence of the appearance of the following two independent effective operators at NLO:
\bea
\cO_{\Omega_3} &=&  {\rm Tr } \left(\hat H_{\mu\nu} \hat H_{\mu\nu}^\dagger  \Omega_{3}  \Omega^\dagger_{3} \right)~, \no \\
\cO_{\Omega_1} &=& {\rm Tr } \left(\hat H_{\mu\nu} \hat H_{\mu\nu}^\dagger  \Omega_{1}  \Omega^\dagger_{1} \right)~, 
\eea
where $\hat H_{\mu\nu}= T_\alpha H^\alpha_{\mu\nu}$. In the custodial limit, they reduce to a single operator, proportional to ${\rm Tr } (\hat H_{\mu\nu} \hat H_{\mu\nu}^\dagger)$, giving rise to a universal field renormalization. Hence, in the $SU(4)_V$ custodial limit, any sensitivity to the radial modes (in the limit of heavy masses) can be completely absorbed by the on-shell renormalization of the (universal) vector coupling $g_4$.

Beyond the custodial limit, we can effectively estimate the error of neglecting the radial modes when employing a universal vector-field renormalization, by looking at the residual non-universal scale dependence in $\Sigma^{(4)}_{V,1}$. For instance, if we renormalize the coupling with an on-shell LQ process, the leading effect induced on the $\Zp$ coupling when neglecting heavy (degenerate) radial modes is proportional to $(z^{\phi}_{\Zp}-z^{\phi}_{U}) \log(M_R^2/m_U^2)$. We can correct for this leading effect, remaining with an ambiguity of the type $(z^{\phi}_{\Zp}-z^{\phi}_{U}) \times\cO(1)$, which is expected to be small.

Finally, we address the question of how large the impact of radial modes can be if they are not heavy. In this case, we do expect finite effects independently of the renormalization procedure. Considering for simplicity the limit where all radial modes have the same mass, $M_R=M_U$, and employing the custodial limit for the vector masses as well, we find 
\begin{align}
\delta \Sigma_U^R (0) = \frac{17 \pi}{6 \sqrt{3}}-5 \approx 0.14\,,
\end{align}
with $\delta \Sigma_U$ as defined in Ref.~\cite{Fuentes-Martin:2019ign}. Even for $g_4=3$, this result leads to a correction to the matching coefficients evaluated in Ref.~\cite{Fuentes-Martin:2019ign} below $1\%$.

For completeness, we also report the corresponding effects at $\cO(\alpha_s)$ in the same mass limit $M_R=M_U$. Employing the same notation as in Section~\ref{sec:nlo}, we obtain
\bea
 \delta_g^{(s,R)} &=&- \frac{2}{3} + \frac{4}{27}\sqrt{3}\pi\, \approx  0.14~,  \no\\
\bar c ^{(s, R)}_{LL}&=& \bar c ^{(s, R)}_{LR} = \bar c ^{(s, R)}_{de} = -\frac{8}{3} + \frac{4}{9}\sqrt{3}\pi  \approx  - 0.25~,
\qquad 
\eea
which are also negligible.

\section{Conclusions}\label{sec:conclusions}

The growing interest on leptoquark-mediated processes, both at low and at high energies, calls for analyses of the corresponding amplitudes beyond tree-level accuracy. In the case of a massive spin-1 leptoquark, this is possible only if the field is embedded into a consistent model with additional degrees of freedom. In this paper, we have clarified two open issues concerning NLO effects in LQ-mediated amplitudes, assuming the LQ to be the massive gauge boson arising from the spontaneous symmetry breaking  
$SU(4) \times  SU(3)^\prime \times U(1)_X \to SU(3)_c \times U(1)_Y$.

On one hand, we have computed the $\cO(\alpha_s)$ corrections to the high-scale matching conditions  of semileptonic operators. These corrections are of the order of a few percent for LQ masses in the few TeV range  and, in the parameter space  of phenomenological interest, are clearly subleading compared to the $\cO(\alpha_4)$ corrections evaluated in Ref.~\cite{Fuentes-Martin:2019ign}. We have clarified two key ingredients that are necessary for a consistent evaluation of the $\cO(\alpha_s)$ corrections:
i)~the identification of all the colored states at the TeV scale;
ii)~the necessity to express at $\cO(\alpha_s)$  accuracy the LQ coupling in terms of high-energy observables.
As we have shown, in the processes considered here, both effects significantly reduce the overall impact of the QCD corrections. 

On the other hand, we have analyzed the dependence of NLO corrections, both at $\cO(\alpha_4)$ and at $\cO(\alpha_s)$ accuracy, from the (unknown) spectrum of massive scalars, which necessarily appear in the sector responsible for the $SU(4) \times  SU(3)^\prime \times U(1)_X \to SU(3)_c \times U(1)_Y$ symmetry breaking. As we have shown, the infinite mass limit for the radial modes can lead to ambiguities in the estimate of NLO effects on LQ-mediated amplitudes if the spectrum of the heavy vectors is not degenerate. However, in the minimal setup, where the symmetry is broken via the scalar fields $\Omega_{1,3}$ and under the assumption of an almost degenerate spectrum of radial modes, we have also shown that these ambiguities are negligible already at $\cO(\alpha_4)$.

\section*{Acknowledgements}

We would like to thank H.~Patel for technical support on Package-X~\cite{Patel:2015tea}.
This project has received funding from the European Research Council (ERC) under the European Union's Horizon 2020 research and innovation programme under grant agreement 833280 (FLAY), and by the Swiss National Science Foundation (SNF) under contract 200021-159720. The work of J.F. was also supported in part by the Generalitat Valenciana under contract SEJI/2018/033.

\appendix

\section{Loop functions}
\label{sect:app}

\subsection{LQ two-point function at \texorpdfstring{$\cO(\alpha_s)$}{O(as)}}

Separating coloron and gluon contributions, the two-point LQ functions at $\cO(\alpha_s)$, in the absence of radial modes, is decomposed as 
\begin{equation}
i\frac{\alpha_s}{4\pi} \left[ \Sigma_U^{(s,\Gp)}(s)+\Sigma_U^{(s,g)}(s)\right]  g^{\mu\nu}~.
\end{equation}
The coloron contribution is 
\bea
&& \Sigma_U^{(s,\Gp)}(s) = \Sigma_{U,0}^{(s,\Gp)} +C_F\left\{ s \left[ 
- \frac{10}{3}\Delta_U - \frac{32}{9} \right. \right.  \no\\
&&\quad +\left. \frac{10x_{G'}}{3} f(x_{G'}) \right] +\left[\frac{6 m_U^4}{s}\left(\frac{{}x_{G'}^2}{9} -\frac{2 x_{G'}}{9}+\frac{1}{9}\right) -\frac{10 s}{3}  \right. \no\\
&&\qquad \left.\left. + m_U^2\left(x_{G'}^2-\frac{13 x_{G'}}{3} -\frac{4}{3}\right)\right] F(s,m_{G'}^2,m_U^2) \right\}~, 
\eea
where 
\bea
\Delta_U &=&  \frac{1}{\epsUV} -\gamma_{\rm E} +\log(4\pi) + \LmuUV~,  \no\\
 f(x) &=& \frac{\log(x)}{x-1}, 
 \eea
and $F(s, m_X^2, m_Y^2)$, defined as in~\cite{Bohm:1986rj}, satisfies  $F(0, m_X^2, m_Y^2)=0$. 
The gluon contribution is 
\bea
&& \Sigma_U^{(s,g)}(s) = \Sigma_{U,0}^{(s,g)} + C_F\left\{ s \left[  \frac{10}{3}\Delta_U +
 \frac{62}{9} \right] - \frac{2 m_{U}^4}{3s}  \right. \no\\ 
 && \left. + 6\left(-\frac{m_{U}^6}{9 s^2}+\frac{m_U^4}{3 s} +\frac{m_U^2}{3}-\frac{5 s}{9}\right) \log\left(1-\frac{s}{m_U^2}\right)\right\}. \no\\
\eea
Finally, the contribution of radial modes, which is finite up to mass correction terms, can be written as 
\be 
\Sigma_U^{(s,R)}(s) = \Sigma_U^{(s,R)} (0) - C_F m_U^2 x_{G'}^2 F(s,m_{G'}^2,m_R^2)~.
\ee

\subsection{LQ vertex correction at \texorpdfstring{$\cO(\alpha_s)$}{O(as)}}

Similarly to the two-point function, we decompose the  $\cO(\alpha_s)$ corrections to the LQ vertex as 
\begin{equation}
i\frac{g_4}{\sqrt{2}}\gamma^\mu \frac{\alpha_s}{4\pi} \left[ \delta V^{(s,\Gp)}(s) + \delta V^{(s,g)}(s)\right]\,.
\end{equation}
The coloron contribution is 
\bea
	\delta V^{(s,\Gp)}(s) &=& C_F\Big[ -3\Delta_U - \frac{5}{2} + 3 x_{G'}  f(x_\Gp)  \no\\ 
	&& - 3 \Lambda_4(s,m_{G'},m_U)\Big]~,
\eea
where  $\Lambda_4(s, m_X, m_Y)$, defined as in~\cite{Bohm:1986rj}, satisfies  $\Lambda_4(0, m_X, m_Y)=0$.
The gluon contribution, for generic values of $s$, can be decomposed as 
\bea
&&	\delta V^{(s,g)}(s) =  C_F \left\{ 3\Delta_U - 2 \left(\frac{1}{\epsIR}+ \LmuIR \right)  \right. \no\\
&& \qquad + \frac{2m_U^2}{s}\log\left(\frac{m_U^2}{m_U^2-s}\right)\left(\frac{1}{\epsIR}+ \LmuIR \right) 	\no\\ 
&& \qquad  -\frac{m_U^2}{s}\left[ 1-\log^2\left(\frac{m_U^2}{m_U^2-s}\right) 
 + 2\text{Li}_2\left(\frac{s}{s-m_U^2}\right)\right] \no\\
 && \qquad  \left.	+ \left(\frac{m_U^4}{s^2}-1\right)\log\left(\frac{m_U^2}{m_U^2-s}\right) \right\}.
\eea

\subsection{Vector two-point functions at \texorpdfstring{$\cO(\alpha_4)$}{O(a4)}}

Following the decomposition in (\ref{eq:twopoint4}), 
we present here the 
explicit expressions of the 
$\Sigma^{(4)}_{V,1}$ coefficients and the $\Sigma^{(4)}_{V,{\rm reg}} (s)$ functions.
The regular terms are defined up to constant and linear terms in $s$ that we can reabsorb 
in $\Sigma^{(4)}_{V,0}$ and $\Sigma^{(4)}_{V,1}$, respectively. 
For simplicity, we define them to be
\begin{widetext}
\bea
\Sigma^{(4)}_{U,{\rm reg}} (s)& =&
\frac{N_f}{3}s \log \left(-\frac{s}{m_U^2}\right) \no \\ 
&&+ \left[\frac{m_U^4}{s}\left(-\frac{x_{G'}^3}{9}-\frac{2x_{G'}^2}{3}+ \frac{5x_{G'}}{3}-\frac{8}{9} \right) -s\left( \frac{x_{G'}}{9}-\frac{40}{9}\right)  +m_{U}^2\left( \frac{16}{9}+6x_{G'}-\frac{10x_{G'}^2}{9}\right) \right] F(s,m_U^2, m_{G'}^2) \no \\
&&+\left[\frac{m_U^4}{s}\left(-\frac{x_{Z'}^3}{18}-\frac{x_{Z'}^2}{3} +\frac{5x_{Z'}}{6}-\frac{4}{9} \right)-s\left( \frac{x_{Z'}}{18}-\frac{20}{9}\right) +m_{U}^2\left( \frac{8}{9} + 3x_{Z'}- \frac{5x_{Z'}^2}{9} \right)  \right]F(s, m_U^2, m_{Z'}^2)~,
\no\\
 \Sigma^{(4)}_{\Zp,{\rm reg}} (s)  & =&  \frac{N_f}{3}s\log\left(-\frac{s}{m_U^2}\right) + \left[s\left(6 + \frac{2 x_{Z'}}{3} -\frac{x_{Z'}^2}{6}\right) + m_U^2 \left(8 + \frac{16 x_{Z'}}{3} -\frac{1}{3} 10 x_{Z'}^2\right) \right]F(s,m_U^2,m_U^2)~, \no\\
		 \Sigma^{(4)}_{\Gp,{\rm reg}} (s) &=&  \frac{N_f}{3}s\log\left(-\frac{s}{m_U^2}\right) + \left[\frac{39 }{8}s + \frac{15}{2} m_U^2 x_{G'} \right]F(s,m_{G'}^2,m_{G'}^2) \no\\
&& +\left[s \left(\frac{3}{2}+\frac{x_{G'}}{6}-\frac{x_{G'}^2}{24}\right) + m_U^2 \left(2+\frac{4}{3}x_{G'}-\frac{5}{6} x_{G'}^2\right)\right]F(s,m_U^2,m_U^2)~.
\eea
With this choice, the $\Sigma^{(4)}_{V,1}$ are
\bea
 \Sigma^{(4)}_{U,1} &=& \left(\frac{20}{3} - \frac{N_f}{3}-\frac{ x_{Z'}}{18} - \frac{x_{G'}}{9}\right)\Delta_U + \left(\frac{64}{9} -\frac{5N_f}{9} - \frac{5 x_{Z'}}{54} - \frac{5x_{G'}}{27}\right) \no \\
 &&- \left(\frac{40}{9} - \frac{x_{G'}}{9}\right) x_{G'} f(x_\Gp) -  \left(\frac{20}{9} - \frac{x_{Z'}}{18}\right) x_{Z'} f(x_\Zp)~, \no\\
 \Sigma^{(4)}_{\Gp,1} &=& \left(\frac{51}{8} - \frac{N_f}{3} + \frac{x_{G'}}{6} -\frac{x_{G'}^2}{24}\right)\Delta_U - \left(\frac{5N_f}{9} - \frac{1}{4} - \frac{x_{G'}}{9} +\frac{x_{G'}^2}{36}\right) -\frac{39}{8}\log\left(x_{G'}\right)~, \no\\
 \Sigma^{(4)}_{\Zp,1} &=& \left(\frac{18}{3} - \frac{N_f}{3} + \frac{2 x_{Z'}}{3} - \frac{x_{Z'}^2}{6} \right)\Delta_U -  \left(\frac{5N_f}{9} - \frac{4 x_{Z'}}{9}  + \frac{x_{Z'}^2}{9}\right)~.
\eea
As expected, the expressions for $\Sigma^{(4)}_{V,1}$ or $\Sigma^{(4)}_{V,{\rm reg}} (s)$ are all equal in the $SU(4)_V$ custodial limit, i.e for $m_U=m_{Z^\prime}=m_{G^\prime}$. Also note that in this limit $\Sigma^{(4)}_{V,{\rm reg}} (0)=0$. 
\end{widetext}

\bibliographystyle{JHEP}
\bibliography{references}

\providecommand{\href}[2]{#2}\begingroup\raggedright\begin{thebibliography}{10}

\bibitem{deSimone:2020kwi}
P.~de~Simone, \emph{{Experimental Review on Lepton Universality and Lepton
  Flavour Violation tests in B decays}},
  \href{https://doi.org/10.1051/epjconf/202023401004}{\emph{EPJ Web Conf.}
  {\bfseries 234} (2020) 01004}.

\bibitem{Pati:1974yy}
J.~C. Pati and A.~Salam, \emph{{Lepton Number as the Fourth Color}},
  \href{https://doi.org/10.1103/PhysRevD.10.275,
  10.1103/PhysRevD.11.703.2}{\emph{Phys. Rev.} {\bfseries D10} (1974)
  275--289}.

\bibitem{Alonso:2015sja}
R.~Alonso, B.~Grinstein and J.~Martin~Camalich, \emph{{Lepton universality
  violation and lepton flavor conservation in $B$-meson decays}},
  \href{https://doi.org/10.1007/JHEP10(2015)184}{\emph{JHEP} {\bfseries 10}
  (2015) 184}, [\href{https://arxiv.org/abs/1505.05164}{{\ttfamily
  1505.05164}}].

\bibitem{Calibbi:2015kma}
L.~Calibbi, A.~Crivellin and T.~Ota, \emph{{Effective Field Theory Approach to
  $b\to s\ell\ell^{(')}$, $B\to K^{(*)}\nu\overline{\nu}$ and $B\to
  D^{(*)}\tau\nu$ with Third Generation Couplings}},
  \href{https://doi.org/10.1103/PhysRevLett.115.181801}{\emph{Phys. Rev. Lett.}
  {\bfseries 115} (2015) 181801},
  [\href{https://arxiv.org/abs/1506.02661}{{\ttfamily 1506.02661}}].

\bibitem{Barbieri:2015yvd}
R.~Barbieri, G.~Isidori, A.~Pattori and F.~Senia, \emph{{Anomalies in
  $B$-decays and $U(2)$ flavour symmetry}},
  \href{https://doi.org/10.1140/epjc/s10052-016-3905-3}{\emph{Eur. Phys. J.}
  {\bfseries C76} (2016) 67},
  [\href{https://arxiv.org/abs/1512.01560}{{\ttfamily 1512.01560}}].

\bibitem{Buttazzo:2017ixm}
D.~Buttazzo, A.~Greljo, G.~Isidori and D.~Marzocca, \emph{{B-physics anomalies:
  a guide to combined explanations}},
  \href{https://doi.org/10.1007/JHEP11(2017)044}{\emph{JHEP} {\bfseries 11}
  (2017) 044}, [\href{https://arxiv.org/abs/1706.07808}{{\ttfamily
  1706.07808}}].

\bibitem{Crivellin:2018yvo}
A.~Crivellin, C.~Greub, D.~Müller and F.~Saturnino, \emph{{Importance of Loop
  Effects in Explaining the Accumulated Evidence for New Physics in B Decays
  with a Vector Leptoquark}},
  \href{https://doi.org/10.1103/PhysRevLett.122.011805}{\emph{Phys. Rev. Lett.}
  {\bfseries 122} (2019) 011805},
  [\href{https://arxiv.org/abs/1807.02068}{{\ttfamily 1807.02068}}].

\bibitem{Barbieri:2016las}
R.~Barbieri, C.~W. Murphy and F.~Senia, \emph{{B-decay Anomalies in a Composite
  Leptoquark Model}},
  \href{https://doi.org/10.1140/epjc/s10052-016-4578-7}{\emph{Eur. Phys. J.}
  {\bfseries C77} (2017) 8},
  [\href{https://arxiv.org/abs/1611.04930}{{\ttfamily 1611.04930}}].

\bibitem{Assad:2017iib}
N.~Assad, B.~Fornal and B.~Grinstein, \emph{{Baryon Number and Lepton
  Universality Violation in Leptoquark and Diquark Models}},
  \href{https://doi.org/10.1016/j.physletb.2017.12.042}{\emph{Phys. Lett.}
  {\bfseries B777} (2018) 324--331},
  [\href{https://arxiv.org/abs/1708.06350}{{\ttfamily 1708.06350}}].

\bibitem{Calibbi:2017qbu}
L.~Calibbi, A.~Crivellin and T.~Li, \emph{{Model of vector leptoquarks in view
  of the $B$-physics anomalies}},
  \href{https://doi.org/10.1103/PhysRevD.98.115002}{\emph{Phys. Rev.}
  {\bfseries D98} (2018) 115002},
  [\href{https://arxiv.org/abs/1709.00692}{{\ttfamily 1709.00692}}].

\bibitem{Barbieri:2017tuq}
R.~Barbieri and A.~Tesi, \emph{{$B$-decay anomalies in Pati-Salam SU(4)}},
  \href{https://doi.org/10.1140/epjc/s10052-018-5680-9}{\emph{Eur. Phys. J.}
  {\bfseries C78} (2018) 193},
  [\href{https://arxiv.org/abs/1712.06844}{{\ttfamily 1712.06844}}].

\bibitem{Blanke:2018sro}
M.~Blanke and A.~Crivellin, \emph{{$B$ Meson Anomalies in a Pati-Salam Model
  within the Randall-Sundrum Background}},
  \href{https://doi.org/10.1103/PhysRevLett.121.011801}{\emph{Phys. Rev. Lett.}
  {\bfseries 121} (2018) 011801},
  [\href{https://arxiv.org/abs/1801.07256}{{\ttfamily 1801.07256}}].

\bibitem{DiLuzio:2017vat}
L.~Di~Luzio, A.~Greljo and M.~Nardecchia, \emph{{Gauge leptoquark as the origin
  of B-physics anomalies}},
  \href{https://doi.org/10.1103/PhysRevD.96.115011}{\emph{Phys. Rev.}
  {\bfseries D96} (2017) 115011},
  [\href{https://arxiv.org/abs/1708.08450}{{\ttfamily 1708.08450}}].

\bibitem{DiLuzio:2018zxy}
L.~Di~Luzio, J.~Fuentes-Martin, A.~Greljo, M.~Nardecchia and S.~Renner,
  \emph{{Maximal Flavour Violation: a Cabibbo mechanism for leptoquarks}},
  \href{https://doi.org/10.1007/JHEP11(2018)081}{\emph{JHEP} {\bfseries 11}
  (2018) 081}, [\href{https://arxiv.org/abs/1808.00942}{{\ttfamily
  1808.00942}}].

\bibitem{Bordone:2017bld}
M.~Bordone, C.~Cornella, J.~Fuentes-Martin and G.~Isidori, \emph{{A three-site
  gauge model for flavor hierarchies and flavor anomalies}},
  \href{https://doi.org/10.1016/j.physletb.2018.02.011}{\emph{Phys. Lett.}
  {\bfseries B779} (2018) 317--323},
  [\href{https://arxiv.org/abs/1712.01368}{{\ttfamily 1712.01368}}].

\bibitem{Bordone:2018nbg}
M.~Bordone, C.~Cornella, J.~Fuentes-Martín and G.~Isidori, \emph{{Low-energy
  signatures of the $\mathrm{PS}^3$ model: from $B$-physics anomalies to LFV}},
  \href{https://doi.org/10.1007/JHEP10(2018)148}{\emph{JHEP} {\bfseries 10}
  (2018) 148}, [\href{https://arxiv.org/abs/1805.09328}{{\ttfamily
  1805.09328}}].

\bibitem{Greljo:2018tuh}
A.~Greljo and B.~A. Stefanek, \emph{{Third family quark–lepton unification at
  the TeV scale}},
  \href{https://doi.org/10.1016/j.physletb.2018.05.033}{\emph{Phys. Lett.}
  {\bfseries B782} (2018) 131--138},
  [\href{https://arxiv.org/abs/1802.04274}{{\ttfamily 1802.04274}}].

\bibitem{Cornella:2019hct}
C.~Cornella, J.~Fuentes-Martin and G.~Isidori, \emph{{Revisiting the vector
  leptoquark explanation of the B-physics anomalies}},
  \href{https://doi.org/10.1007/JHEP07(2019)168}{\emph{JHEP} {\bfseries 07}
  (2019) 168}, [\href{https://arxiv.org/abs/1903.11517}{{\ttfamily
  1903.11517}}].

\bibitem{Fuentes-Martin:2020bnh}
J.~Fuentes-Martín and P.~Stangl, \emph{{Third-family quark-lepton unification
  with a fundamental composite Higgs}},
  \href{https://arxiv.org/abs/2004.11376}{{\ttfamily 2004.11376}}.

\bibitem{Guadagnoli:2020tlx}
D.~Guadagnoli, M.~Reboud and P.~Stangl, \emph{{The Dark Side of 4321}},
  \href{https://arxiv.org/abs/2005.10117}{{\ttfamily 2005.10117}}.

\bibitem{Georgi:2016xhm}
H.~Georgi and Y.~Nakai, \emph{{Diphoton resonance from a new strong force}},
  \href{https://doi.org/10.1103/PhysRevD.94.075005}{\emph{Phys. Rev.}
  {\bfseries D94} (2016) 075005},
  [\href{https://arxiv.org/abs/1606.05865}{{\ttfamily 1606.05865}}].

\bibitem{Diaz:2017lit}
B.~Diaz, M.~Schmaltz and Y.-M. Zhong, \emph{{The leptoquark Hunter’s guide:
  Pair production}}, \href{https://doi.org/10.1007/JHEP10(2017)097}{\emph{JHEP}
  {\bfseries 10} (2017) 097},
  [\href{https://arxiv.org/abs/1706.05033}{{\ttfamily 1706.05033}}].

\bibitem{Fuentes-Martin:2019ign}
J.~Fuentes-Martín, G.~Isidori, M.~König and N.~Selimović, \emph{{Vector
  Leptoquarks Beyond Tree Level}},
  \href{https://doi.org/10.1103/PhysRevD.101.035024}{\emph{Phys. Rev.}
  {\bfseries D101} (2020) 035024},
  [\href{https://arxiv.org/abs/1910.13474}{{\ttfamily 1910.13474}}].

\bibitem{Aebischer:2018acj}
J.~Aebischer, A.~Crivellin and C.~Greub, \emph{{QCD improved matching for
  semileptonic $B$ decays with leptoquarks}},
  \href{https://doi.org/10.1103/PhysRevD.99.055002}{\emph{Phys. Rev.}
  {\bfseries D99} (2019) 055002},
  [\href{https://arxiv.org/abs/1811.08907}{{\ttfamily 1811.08907}}].

\bibitem{Baker:2019sli}
M.~J. Baker, J.~Fuentes-Martín, G.~Isidori and M.~König, \emph{{High- $p_T$
  signatures in vector–leptoquark models}},
  \href{https://doi.org/10.1140/epjc/s10052-019-6853-x}{\emph{Eur. Phys. J.}
  {\bfseries C79} (2019) 334},
  [\href{https://arxiv.org/abs/1901.10480}{{\ttfamily 1901.10480}}].

\bibitem{Patel:2015tea}
H.~H. Patel, \emph{{Package-X: A Mathematica package for the analytic
  calculation of one-loop integrals}},
  \href{https://doi.org/10.1016/j.cpc.2015.08.017}{\emph{Comput. Phys. Commun.}
  {\bfseries 197} (2015) 276--290},
  [\href{https://arxiv.org/abs/1503.01469}{{\ttfamily 1503.01469}}].

\bibitem{Bohm:1986rj}
M.~Bohm, H.~Spiesberger and W.~Hollik, \emph{{On the One Loop Renormalization
  of the Electroweak Standard Model and Its Application to Leptonic
  Processes}}, \href{https://doi.org/10.1002/prop.19860341102}{\emph{Fortsch.
  Phys.} {\bfseries 34} (1986) 687--751}.

\end{thebibliography}\endgroup

\end{document}